\documentclass[a4paper,12pt]{article}

\usepackage{amsmath}
\usepackage{amssymb}
\usepackage{amsthm}

\newcommand{\reals}{\ensuremath{\mathbb{R}}}
\newcommand{\prt}{\partial}
\newcommand{\vertbar}{{\bigg\arrowvert}}

\def\Pr{{\mathbb P}}

\def\B{\mathcal{B}}
\def\V{\mathcal{V}}
\begin{document}
\bibliographystyle{plain}

\title{{\Large\bf The correspondence between Tracy-Widom (TW) and Adler-Shiota-van Moerbeke (ASvM) approaches in random matrix theory: the Gaussian case}}
\author{Igor Rumanov\footnote{e-mail: igorrumanov@math.ucdavis.edu} \\
{\small Department of Mathematics, UC Davis, 1 Shields Avenue, CA 95616}}

\maketitle

%$\vspace{2cm}$
\bigskip
%$\hspace{5cm}$ {\large\bf Abstract} \\
\begin{abstract}
Two approaches (TW and ASvM) to derivation of integrable differential equations for random matrix probabilities are compared. Both methods are rewritten in such a form that simple and explicit relations between all TW dependent variables and $\tau$-functions of ASvM are found, for the example of finite size Gaussian matrices. Orthogonal function systems  and Toda lattice are seen as the core structure of both approaches and their relationship.
\end{abstract}

\newpage

\section*{\normalsize\bf I. INTRODUCTION}

\par Random matrices were introduced in the middle of last century by Wigner and Dyson as a description of complex systems that were difficult to study since they did not admit any exact or even approximate solutions. So the fact, thoroughly realized in the early nineties, is still amazing: many interesting random matrix (RM) models are closely related with and conveniently described by integrable systems --- integrable hierarchies of partial differential equations (PDE). The first such result by Jimbo, Miwa, M\^ori and Sato$^{13}$ appeared earlier but remained unrecognized as an instance of a general rule for a decade. Matrix integrals, giving the probability distributions for eigenvalues of random matrices, proved to be $\tau$-functions of integrable hierarchies. The ground-breaking works of Tracy and Widom$^{16,17}$ followed, which generalized and extended the results of Ref. $13$ for the sine kernel to a number of other cases. The authors derived integrable PDE or Painlev\'e ODE as equations satisfied by the probabilities for the spectrum of various random matrices as well as by their large size asymptotics.
\par Another major approach was developed by Adler, Shiota and van Moerbeke$^{1-6}$. It is based directly on the correspondence between matrix integrals over the spectral domains and $\tau$-functions --- the ``generating functions" of integrable PDE. Using both methods, different integrable PDE have been obtained not only for single matrix ensembles but also for coupled Hermitian matrices$^{4,5,8,18}$. In the latter case equations derived by different methods seem to be so dissimilar that the only indication of the existence of relations between them is that they are equations for the same probabilities.  
\par In this paper, we compare the ``algebraic" approach of Adler, Shiota and van Moerbeke with the ``functional-theoretic" approach of Tracy and Widom and show the common structure behind them. This structure is (not surprisingly) the Toda lattice$^{19,1,3}$. The relation between these two approaches has not been heretofore completely clarified. The reason lies in the fact that the authors of Refs. $2,3$ used the KP hierarchy rather than one-dimensional Toda lattice hierarchy (1-Toda) to derive their PDE for single Hermitian matrices. This led to equations very different in form from those derived in Refs. $16,17$. J.Harnad$^{11}$ demonstrated for the special cases of Airy and Bessel kernels, how the ``KP"-equation of ASvM can be obtained by taking the appropriate combinations of TW equations. We show that there is a more direct and complete correspondence of structures involving the Toda lattice hierarchy; and thus the procedure of Ref. $11$ is essentially an implicit transition from 1-Toda to KP form of integrable equations related to Hermitian 1-matrix models. In this paper we restrict ourselves to the Hermitian one-matrix Gaussian case. Our main purpose is to reveal explicitly the underlying structures connecting dependent variables and equations in the different approaches. As we show by this example, for single matrices equations obtained by different methods (Ref. $17$ vs. Refs. $2,3$) can essentially coincide, despite the very different ways to arise. For comparison, we give the implementation of the Harnad's approach in the Appendix for the case of finite size Gaussian matrices considered here.
\par In another related work$^{10}$, 1-Toda lattice was ingeniously used to treat the Virasoro constraints of ASvM approach, which, together with an equation of Ref. $17$, helped to find Painlev\'e VI equation for the Jacobi ensemble.

\section*{\normalsize\bf II. SUMMARY OF RESULTS}

\par We use the ASvM approach of Hirota's bilinear identities for $\tau$-functions of the $1$-dimensional Toda lattice hierarchy$^{19}$ and (linear) Virasoro constraints. Here the $\tau$-function $\tau_n^J$ has the meaning of complex Hermitian Gaussian matrix integral over spectral domain $J^n$, $J = \cup_i (a_{2i-1}, a_{2i})$ -- a finite union of intervals of the real axis $\reals$ with endpoints $a_i$:

$$
\tau_n^J(t) = \frac{1}{n!}\int_{J^n} \prod_{i<j} (x_i-x_j)^2 \prod_1^n \rho_t(x_i) dx_i,
$$

\noindent where $n$ is the random matrix size, $t = (t_1, t_2, t_3, ...)$ is a complex infinite vector, $\rho_t(x) = e^{-V(x)+\sum_{k=1}^{\infty}t_kx^k}$ -- the $t$-deformed weight; and $\tau_n(t)$ will denote the corresponding integral over the whole $\reals^n$. Our first results are the equations

$$
\B_{-1}^2\ln \tau_n^J = 2(2Q_nP_n - n), \hspace{3cm} \eqno(1)%(0)
$$

$$
\B_{-1}^2 Q_n = (-2\B_0 + 4n)Q_n - 8Q_n^2P_n, \hspace{3cm} \eqno(2)%(+)
$$

$$
\B_{-1}^2 P_n = (2\B_0 + 4n)P_n - 8Q_nP_n^2, \hspace{3cm} \eqno(3)%(-)
$$

\noindent where $Q_n = \tau_{n+1}^J/\tau_n^J$, $P_n = \tau_{n-1}^J/\tau_n^J$, and the operators $\B_k$ contain partial derivatives w.r.t. the boundary points of the intervals:

$$
\B_k\tau_n^J = \sum_{i \in \{endpoints\}} a_i^{k+1}\frac{\partial}{\partial a_i}\tau_n^J.
$$

Then, analyzing the TW equations for the finite $n$ Hermite case, we obtain simple relations among the above $\tau$-functions and their ratios and the auxiliary functions $u, v, w$, appearing in Ref. $17$. The TW approach consists in studying the kernel of a Fredholm integral operator $K_n^J$, whose determinant is the probability of having no eigenvalues in a subset of $\reals$, here -- the complement $J^c$ of the set $J$:

$$
\det(I - K_n^J) = \frac{\tau^J_n}{\tau_n}.\ \eqno(4)%(1.1)
$$

\noindent where the numerator and denominator of the r.h.s. are the matrix integrals (the $\tau$ functions $\tau(t=0)$) over $J$ and over real axis $\reals$, respectively. The operator $K_n^J$ has kernel

$$
K_n^J(x,y) = K_n(x, y)\chi_{J^c}(y),\ \eqno(5)%(1.2)
$$

\noindent where $\chi_{J^c}(x)$ is the characteristic function of $J^c$:

$$
\chi_{J^c}(x) = \left\{\begin{array}{cc} 1,\ & if\ x \in J^c \\ 0,\ & if\ x \notin J^c \end{array}\right.,
$$

\noindent $K_n(x, y)$ is an integrable kernel$^{12}$, which can be expressed by Christoffel-Darboux formula for orthogonal functions,

$$
K_n(x, y) = \frac{\phi(x)\psi(y) - \psi(x)\phi(y)}{x - y} = \sum_{k=0}^{n-1}\phi_k(x)\phi_k(y),\ \eqno(6)%(1.3)
$$

\noindent $\phi(x) = \sqrt{b_{n-1}}\phi_n(x)$, $\psi(x) = \sqrt{b_{n-1}}\phi_{n-1}(x)$, and $b_{n-1}$ is the coefficient of $\phi_{n-1}(x)$ in the corresponding 3-term recurrence relation.
\par We show that the inner product $v = (\phi, (I-K_n^J)^{-1}\psi)$ is related to $\tau_n^J$ in the following way:

$$
v \equiv (\phi, (I-K_n^J)^{-1}\psi) = \frac{\B_{-1}\ln\tau_n^J}{2}.\ \eqno(7)%(*0)
$$

\noindent Also, if we denote 

$$
\tilde Q_n = \sqrt{n/2}-u,\hspace{1cm} \tilde P_n = w+\sqrt{n/2},\ \eqno(8)%(1.4)
$$

\noindent where $u = (\phi, (I-K_n^J)^{-1}\phi)$, $w = (\psi, (I-K_n^J)^{-1}\psi)$, then these functions $\tilde Q_n$ and $\tilde P_n$ turn out to satisfy the same nonlinear PDE (2) and (3) as the ratios of $\tau$-functions $Q_n$ and $P_n$, respectively. These facts lead to even stronger conclusions:

$$
\tilde Q_n = C_nQ_n, \hspace{1cm} \tilde P_n = Q_n/C_n,%\ \eqno(1.5)
$$

\noindent where $C_n = (2^n/n!)\cdot\sqrt{n/2\pi}$. So the functions $u$ and $w$ appearing in the resolvent kernel approach, are represented in terms of ratios of matrix integrals ($\tau$-functions at $t=0$):

$$
u \equiv (\phi, (I-K_n^J)^{-1}\phi) = \sqrt\frac{n}{2}\left(1-\frac{2^n}{n!\sqrt\pi}\frac{\tau_{n+1}^J}{\tau_n^J}\right) = \sqrt\frac{\tau_{n+1}\tau_{n-1}}{(\tau_n)^2}\left(1-\frac{\tau_{n+1}^J/\tau_{n+1}}{\tau_n^J/\tau_n}\right),\ \eqno(9)%(*+)
$$

$$
w \equiv (\psi, (I-K_n^J)^{-1}\psi) = \sqrt\frac{n}{2}\left(\frac{(n-1)!\sqrt\pi}{2^{n-1}}\frac{\tau_{n-1}^J}{\tau_n^J} - 1\right) = \sqrt\frac{\tau_{n+1}\tau_{n-1}}{(\tau_n)^2}\left(\frac{\tau_{n-1}^J/\tau_{n-1}}{\tau_n^J/\tau_n}-1\right).\ \eqno(10)%(*-)
$$

\par Thus, the correspondence between TW auxiliary variables and $\tau$-functions of one-dimensional Toda lattice hierarchy considered by Adler and van Moerbeke,$^{1,3}$ becomes direct and explicit. The {\it probabilities}, expressed in the form of $\tau$-ratios are seen from the above formulas to be also the natural and convenient auxiliary coordinates. Indeed, $\tau$-functions and their ratios seem to provide the best description for all integrable problems. Here, in the Gaussian case, for instance, we have the following recurrence relations for the $\tau$-ratios:

$$
P_{n+1} = \frac{1}{Q_n},\ \eqno(11)
$$

$$
Q_{n+1} = Q_n\left ( \frac{\B_{-1}^2 \ln Q_n}{4} + Q_nP_n + \frac{1}{2} \right ),\ \eqno(12)
$$

\noindent the first --- by definition of $Q_n$, $P_n$, the second follows from Eq. (1) and the first one. Thus in principle these ratios can be recursively determined from the simplest ones ($Q_0$ and $Q_1$ here, with our normalization, $Q_0 = \tau_1^J = \int_Je^{-x^2}dx$, $Q_1 = \tau_2^J/\tau_1^J$, $\tau_2^J = 1/2\int_{J^2}e^{-x^2-y^2}(x-y)^2dxdy$).
\par {\it Remark.} The 1-Toda lattice $\tau$-functions are naturally related to the orthogonal polynomial structure and this is the reason for the existence of such simple relations. The orthogonal functions for $t$-deformed weight satisfy the $t$-dependent three-term recurrence relation

$$
x\phi_n(x;t) = a_n(t)\phi_n(x;t) + b_n(t)\phi_{n+1}(x;t) + b_{n-1}(t)\phi_{n-1}(x;t),
$$

\noindent where the coefficients depend on times and are known in terms of the Toda $\tau$-functions$^{1,3}$:

$$
a_n(t) = \frac{\prt}{\prt t_1}\ln\frac{\tau_{n+1}}{\tau_n},
$$

$$
b_{n-1}^2(t) = \frac{\tau_{n+1}\tau_{n-1}}{\tau_n^2}.
$$

\noindent These relations show that the 1-Toda time flows are in fact flows between different {\it orthogonal function systems}. The 1-Toda bilinear relations may also be considered as a form of orthogonality relations for systems of orthogonal functions. Recall also that Christoffel-Darboux formula, which is the starting point of the resolvent kernel approach, is itself equivalent to the three-term relations for the corresponding orthogonal functions. One can see that the structure of orthogonal function systems is at the heart of all integrability derivations and connections with random matrix theory.

\section*{\normalsize\bf III. DERIVATION OF EQUATIONS (1),(2),(3): BILINEAR IDENTITIES AND VIRASORO CONSTRAINTS}

\par We consider Hermitian random matrices of size $n \times n$ with spectral measure $d\mu = \rho(x)dx$, $\rho(x) = e^{-V(x)}$ such that all its moments exist. Then the probability that all eigenvalues are in $J \subseteq \reals$ is

$$
\Pr(all\ x_i \in J) = \frac{\int_{J^n} \Delta_n^2(x) \prod_1^n \rho(x_i) dx_i}{\int_{\reals^n} \Delta_n^2(x) \prod_1^n \rho(x_i) dx_i},\ \eqno(13)
$$

\noindent where $\Delta_n(x) = \prod_{i<j} (x_i-x_j)$ is the Vandermonde determinant. The $\tau$-function of integrable hierarchy appears from deformation of the above integrals: let $t = (t_1, t_2, t_3, ...)$ be a complex infinite vector of ``times" and $\rho_t(x) = e^{-V(x)+\sum_{k=1}^{\infty}t_kx^k}$ -- the deformed weight, then the matrix integral

$$
\tau_n^J(t) = \frac{1}{n!}\int_{J^n} \Delta_n^2(x) \prod_1^n \rho_t(x_i) dx_i,\ \eqno(14)%(3.1)
$$

\noindent is a $\tau$-function of integrable hierarchies such as KP and one-dimensional Toda lattice hierarchy (see, for instance, Refs. $1-3$).

\par Matrix integrals for coupled random matrices satisfy$^{4}$ Hirota bilinear identities for 2-dimensional Toda lattice (2-Toda)$^{19}$:

$$
\oint_{z=\infty}\tau_n(t-[z^{-1}], s)\tau_{m+1}(t'+[z^{-1}], s')e^{\sum_1^{\infty}(t_k-t'_k)z^k} z^{n-m-1}dz = 
$$

$$
= \oint_{z=0}\tau_{n+1}(t, s-[z])\tau_m(t', s'+[z])e^{\sum_1^{\infty}(s_k-s'_k)z^{-k}} z^{n-m-1}dz \hspace{1in} \eqno(15)%(2T)
$$

\noindent The 2-Toda $\tau$-functions have two infinite sets of time parameters. There are two possible reductions of them to the bilinear identities for 1-matrix integral $\tau$-functions, possessing only one set of times (first was written in Ref. $3$, second -- in Ref. $6$). The first one can be obtained by setting $\tau(t, s) = \tau(t-s)$ and $s-s'=-(t-t')$, then renaming variables $t-s \to t$, $t'-s' \to t'$. It gives the bilinear identities$^{3}$ for 1-dimensional Toda lattice hierarchy (1-Toda) (including the AKNS hierarchy and the one-dimensional Toda chain itself, see, e.g., Ref. $19$):

$$
\oint_{z=\infty}\tau_n(t-[z^{-1}])\tau_{m+1}(t'+[z^{-1}])e^{1/2\sum_1^{\infty}(t_k-t'_k)z^k} z^{n-m-1}dz = 
$$

$$
= \oint_{z=0}\tau_{n+1}(t+[z])\tau_m(t'-[z])e^{-1/2\sum_1^{\infty}(t_k-t'_k)z^{-k}} z^{n-m-1}dz \hspace{1in} \eqno(16)%(1T)
$$

\noindent The second one is obtained by setting $\tau(t, s) = \tau(t-s)$ and $s=s'$, then again renaming $t-s \to t$, $t'-s' \to t'$. This gives the bilinear identities$^{6}$ for the {\it discrete} KP hierarchy for $n>m$:

$$
\oint_{z=\infty}\tau_n(t-[z^{-1}])\tau_{m+1}(t'+[z^{-1}])e^{\sum_1^{\infty}(t_k-t'_k)z^k} z^{n-m-1}dz = 
$$

$$
= \oint_{z=0}\tau_{n+1}(t+[z])\tau_m(t'-[z]) z^{n-m-1}dz = 0\ for\ n>m \hspace{1in} \eqno(17)%(dKP)
$$

\noindent The most familiar bilinear identity of the usual KP hierarchy$^{9}$ corresponds to taking $m=n-1$ in (17). The integrable nonlinear PDE can be readily obtained from the above identities by shifting times $t\to t+a$, $t'\to t-a$, expanding in Taylor series in $a$ and taking residues for each term of this expansion independently.
\par Let us proceed in the 1-Toda setting, which, as we will see, shows most clearly the correspondence between the ASvM and the TW approaches to the derivation of nonlinear integrable PDE for the spectral probabilities of random matrices.
\par The simplest and most important equations are obtained from (16) for $m=n$ and $m=n-1$ upon taking the residues of the terms linear in $2a = t-t'$. For $m=n$ one gets the hierarchy of equations in terms of $\tau$-functions

$$
\frac{\partial}{\partial t_k}\ln\frac{\tau_{n+1}}{\tau_n} = \sum_{i=0}^k \frac{p_i(-\tilde\partial_t)\tau_n\cdot p_{k-i}(\tilde\partial_t)\tau_{n+1}}{\tau_n\tau_{n+1}}, \hspace{2cm} \eqno(18)%(3.2)
$$

\noindent where $p_k(t)$ are the elementary Schur polynomials defined by

$$
e^{\sum_1^{\infty}t_k z^k} = \sum_{k=0}^{\infty}p_k(t)z^k, \hspace{2cm} \eqno(19)%(3.3)
$$

\noindent $\tilde\partial_t = (\partial_{t_1}, 1/2\partial_{t_2}, 1/3\partial_{t_3}, ...$, so that one can write

$$
\tau_n(t \mp [z^{-1}]) = \sum_{k=0}^{\infty}z^{-k}p_k(\mp\tilde\partial_t)\tau_n(t).\ \eqno(20)
$$

\noindent Equation (18) is trivial when $k=1$. The simplest nontrivial equation arises from (18) when $k=2$. It can be written as

$$
\frac{\partial}{\partial t_2}\ln\frac{\tau_{n+1}}{\tau_n} = \frac{\partial^2}{\partial t_1^2}\ln\frac{\tau_{n+1}}{\tau_n} + \left(\frac{\partial}{\partial t_1}\ln\frac{\tau_{n+1}}{\tau_n}\right)^2 + 2\frac{\partial^2\ln\tau_n}{\partial t_1^2}, \hspace{2cm} \eqno(21a)%(3.4a)
$$

\noindent or, introducing the new variable $Q_n = \tau_{n+1}/\tau_n$, as

$$
\frac{\partial Q_n}{\partial t_2} = \frac{\partial^2 Q_n}{\partial t_1^2} + 2\frac{\partial^2\ln\tau_n}{\partial t_1^2} Q_n. \hspace{2cm} \eqno(21b)%(3.4b)
$$

Now consider the 1-Toda bilinear identity for $m = n-1$. The same procedure of taking residue for terms of order $a$ as above now gives another infinite set of integrable PDE:

$$
2\frac{\partial^2}{\partial t_1\partial t_k}\ln\tau_n - \sum_{i=0}^{k+1} \frac{p_i(-\tilde\partial_t)\tau_n\cdot p_{k+1-i}(\tilde\partial_t)\tau_n}{\tau_n^2} = \sum_{i=0}^{k-1} \frac{p_i(-\tilde\partial_t)\tau_{n+1}\cdot p_{k-1-i}(\tilde\partial_t)\tau_{n-1}}{\tau_n^2}.
$$

\noindent Then the simplest nontrivial equation is obtained when $k=1$, and this is the standard Toda equation for 1-dimensional Toda chain expressed in terms of $\tau$-functions:

$$
\frac{\partial^2\ln\tau_n}{\partial t_1^2} = \frac{\tau_{n+1}\tau_{n-1}}{\tau_n^2}. \hspace{2cm} \eqno(22)%(3-5)
$$

\noindent Thus we obtained a closed system of three nonlinear integrable equations for three dependent variables, $\ln\tau_n$, $Q_n$ and $P_n$: 

$$
\frac{\partial^2\ln\tau_n}{\partial t_1^2} = Q_nP_n, \hspace{4cm} \eqno(23)%(0)
$$

$$
\frac{\partial Q_n}{\partial t_2} = \frac{\partial^2 Q_n}{\partial t_1^2} + 2\frac{\partial^2\ln\tau_n}{\partial t_1^2}Q_n, \hspace{2cm} \eqno(24)%(+)
$$

$$
-\frac{\partial P_n}{\partial t_2} = \frac{\partial^2 P_n}{\partial t_1^2} + 2\frac{\partial^2\ln\tau_n}{\partial t_1^2}P_n, \hspace{2cm} \eqno(25)%(-)
$$

\noindent which is nothing but the well-known Toda-AKNS system (see, for instance, Ref. $15$).
\par Next we will use the well-known Virasoro constraints to express the time derivatives in the previous equations in terms of {\it boundary} derivatives, i.e. the derivatives with respect to the endpoints of the intervals where the matrix eigenvalues lie. To make the paper self-contained, we reproduce here the derivation of Virasoro constraints for the general $\beta$ one-matrix ensemble.$^{7,4,3}$ Although for the purposes of current paper we will need only the simplest case $\beta = 2$ corresponding to Hermitian ensemble, it is instructive and not more difficult to give the derivation for arbitrary real nonnegative $\beta$. The Virasoro constraints arise from the obvious invariance of the matrix integral (the $\tau$-function) under an arbitrary change of integration variables. Consider infinitesimal such changes of the form $x_i \to x_i + \varepsilon x_i^{k+1}$, $\varepsilon \to 0$, $k \ge -1$. Then the generators $\V_k$ of such changes of coordinates make a subalgebra of Virasoro algebra. The change in the integrand, which will be expressed in terms of the time derivatives of the integral, is compensated by the corresponding change of the limits of integration, which in turn will be expressed in terms of the boundary derivatives of the integral. Consider the change in the integrand $dI_n$ of the matrix integral,$^{7,4,3}$

$$
\frac{d(dI_n(x + \varepsilon x^{k+1}))}{d\varepsilon}\vertbar_{\varepsilon = 0},\hspace{1cm} \int_{J^n}dI_n(x) = \tau_n^J(t) = \frac{1}{n!}\int_{J^n} \Delta_n^{\beta}(x) \prod_1^n \rho_t(x_i) dx_i.\ \eqno(14')%(3-1')
$$

\noindent If $k > 0$ the change of the Vandermonde determinant $\Delta_n$ is

$$
\frac{d\Delta_n^{\beta}(x + \varepsilon x^{k+1})}{d\varepsilon}\vertbar_{\varepsilon = 0} = \beta\Delta_n^{\beta-1}(x)\frac{d\Delta_n(x + \varepsilon x^{k+1})}{d\varepsilon}\vertbar_{\varepsilon = 0} = \beta\Delta_n^{\beta}(x)\sum_{i=1}^n\sum_{j>i}^n\frac{x_i^{k+1}-x_j^{k+1}}{x_i-x_j} =
$$

$$
= \Delta_n^{\beta}(x)\left(\frac{1}{2}\sum_{k_1+k_2=k, k_l>0}\sum_{i,j=1}^n x_i^{k_1}x_j^{k_2} + \left(n - \frac{k+1}{2}\right)\sum_{i=1}^n x_i^k\right),
$$

\noindent also

$$
\prod_{i=1}^n\frac{d(d(x_i + \varepsilon x_i^{k+1}))}{d\varepsilon}\vertbar_{\varepsilon = 0} = (k+1)\sum_{i=1}^n x_i^k\cdot\prod_{i+1}^n dx_i.
$$

\noindent Using the identity

$$
\frac{\prt}{\prt t_k} dI_n = \sum_{i=1}^n x_i^k\cdot dI_n,\ \eqno(26)
$$

\noindent one gets (the second term on the r.h.s. is absent for $k=1$)

$$
\frac{d(dI_n(x + \varepsilon x^{k+1}))}{d\varepsilon}\vertbar_{\varepsilon = 0} = 
$$

$$
= \left\{-\sum_{i=1}^n x_i^{k+1}\frac{dV}{dx_i} + \frac{\beta}{2}\sum_{l=1}^{k-1}\frac{\prt^2}{\prt t_l\prt t_{k-l}} + \sum_{l=1}^{\infty}lt_l\frac{\prt}{\prt t_{l+k}} + [\beta n + (1-\beta/2)(k+1)]\frac{\prt}{\prt t_k}\right\}dI_n.
$$

For $k=-1$ the measure $\Delta_n^{\beta}(x) \prod_1^n dx_i$ does not change under translation $x \to x+\varepsilon$, so one has instead

$$
\frac{d(dI_n(x + \varepsilon))}{d\varepsilon}\vertbar_{\varepsilon = 0} = \left\{-\sum_{i=1}^n\frac{dV}{dx_i} + nt_1 + \sum_{l=2}^{\infty}lt_l\frac{\prt}{\prt t_{l-1}}\right\}dI_n.
$$

For $k=0$ one has for the change of the measure

$$
\frac{d(\Delta_n^{\beta}\prod_idx_i(x + \varepsilon x))}{d\varepsilon}\vertbar_{\varepsilon = 0} = \left(\frac{\beta}{2}n(n-1) + n\right)\Delta_n^{\beta}(x)\prod_idx_i,
$$

\noindent and one gets

$$
\frac{d(dI_n(x + \varepsilon x))}{d\varepsilon}\vertbar_{\varepsilon = 0} = \left\{-\sum_{i=1}^n x_i\frac{dV}{dx_i} + \sum_{l=1}^{\infty}lt_l\frac{\prt}{\prt t_l} + n + \frac{\beta}{2}n(n-1)\right\}dI_n.
$$

\par Thus one finally obtains the $k=-1$ constraint:

$$
\B_{-1}\tau_n^J = \V_{-1}\tau_n^J = \int\left\{-\sum_{i=1}^n\frac{dV}{dx_i} + nt_1 + \sum_{l=2}^{\infty}lt_l\frac{\partial}{\partial t_{l-1}}\right\} dI_n(x),\ \eqno(27a)%(3-6a)
$$

\noindent the $k=0$ constraint:

$$
\B_0\tau_n^J = \V_0\tau_n^J = \int\left\{-\sum_{i=1}^nx_i\frac{dV}{dx_i} + n + \beta\frac{n(n-1)}{2} + \sum_{l=1}^{\infty}lt_l\frac{\partial}{\partial t_l}\right\} dI_n(x),\ \eqno(27b)%(3-6b)
$$

\noindent and the $k>0$ constraints (again the second term on the r.h.s. is absent for $k=1$):

$$
\B_k\tau_n^J = \V_k\tau_n^J = 
$$

$$
\int\left\{-\sum_{i=1}^nx_i^{k+1}\frac{dV}{dx_i} + \frac{\beta}{2}\sum_{l=1}^{k-1}\frac{\partial^2}{\partial t_l\partial t_{k-l}} + [\beta n + (1-\beta/2)(k+1)]\frac{\partial}{\partial t_k} + \sum_{l=1}^{\infty}lt_l\frac{\partial}{\partial t_{l+k}}\right\} dI_n(x).\ \eqno(27c)%(3-6c)
$$

\noindent If the potential $V(x)$ is polynomial, $\frac{dV}{dx} = \sum_{l=0}g_lx^l$, then one has

$$
\sum_{i=1}^nx_i^{k+1}\frac{dV}{dx_i} = \sum_{l=0}g_l\frac{\partial}{\partial t_{l+k+1}}.
$$

\noindent The infinitesimal changes of integration limits are expressed in terms of boundary operators$^{1-3}$

$$
\B_k\tau_n^J = \sum_{i \in \{endpoints\}} a_i^{k+1}\frac{\partial}{\partial a_i}\tau_n^J.\ \eqno(28)%(3-7)
$$

\noindent In general, also the next important consequences of the above Virasoro constraints are needed$^{1-3}$:

$$
\B_k\B_l\tau_n^J = \B_k\V_l\tau_n^J = \V_l\B_k\tau_n^J = \V_l\V_k\tau_n^J,\ \eqno(29)%(3-8)
$$

\noindent the second equality holds because the operators $\B_k$ and $\V_l$ commute.
\par From now on we consider the special case of matrices with Gaussian distribution of matrix elements, i.e. Gaussian potential $V(x) = x^2$. At the locus $t_1 = t_2 = ... = 0$, the Virasoro constraints (27a), (27b) together with identity (26) give in this case$^{2,3}$:

$$
\frac{\partial \ln\tau_n^J}{\partial t_1} = -\frac{\B_{-1}\ln \tau_n^J}{2},\ \eqno(30)%(3-9)
$$

$$
\frac{\partial^2 \ln\tau_n^J}{\partial t_1^2} = \frac{\B_{-1}^2\ln \tau_n^J}{4} + \frac{n}{2},\ \eqno(31)%(3-10)
$$

$$
\frac{\partial \ln\tau_n^J}{\partial t_2} = -\frac{\B_0\ln \tau_n^J}{2} + \frac{n^2}{2}.\ \eqno(32)%(3-11)
$$

\noindent These are the only Virasoro constraints we will need here. More such relations (involving also the use of (27c) for $k=1$) are needed for the case of KP-hierarchy,$^{2,3}$ and the derivation is longer. We substitute the relations (30)--(32) into (23), (24), (25) and after a simple algebra obtain the ``boundary form" of equations (23)--(25), which is our system (1)--(3).

%$$
%\B_{-1}^2\ln \tau_n^J = 2(2Q_nP_n - n), \hspace{3cm} \eqno(1)%(0)
%$$

%$$
%\B_{-1}^2 Q_n = (-2\B_0 + 4n)Q_n - 8Q_n^2P_n, \hspace{3cm} \eqno(2)%(+)
%$$

%$$
%\B_{-1}^2 P_n = (2\B_0 + 4n)P_n - 8Q_nP_n^2. \hspace{3cm} \eqno(3)%(-)
%$$

\par If we used the KP hierarchy to derive the PDE as Adler, Shiota and Van Moerbeke did in Refs. $2,3$, rather than 1-Toda, we would have used just one standard integrable PDE --- the KP equation,

$$
\frac{\prt^4 \ln\tau_n}{\prt t_1^4} + 3\frac{\prt^2 \ln\tau_n}{\prt t_2^2} - 4\frac{\prt^2 \ln\tau_n}{\prt t_1\prt t_3} + 6\left(\frac{\prt^2 \ln\tau_n}{\prt t_1^2}\right)^2 = 0.\ \eqno(33)
$$

\noindent Then we would need more Virasoro constraints like eqs. (30)--(32) (involving also the third time $t_3$, the second derivative w.r.t. $t_2$ and the fourth derivative w.r.t. $t_1$). Then the ASvM boundary KP equation would come out, for the finite $n$ Gaussian case it reads$^{2}$:

$$
(\B_{-1}^4 + 8n\B_{-1}^2 + 12\B_{0}^2 + 24\B_{0} - 16\B_{-1}\B_{1})\ln\tau_n^J + 6(\B_{-1}^2\ln\tau_n^J)^2 = 0.\ \eqno(34)%(3-12)
$$

\section*{\normalsize\bf IV. TRACY-WIDOM EQUATIONS AND $\tau$-FUNCTIONS: THE GAUSSIAN CASE}

\par We consider the same Hermitian Gaussian random matrices as above. Various large $n$ limits for this case belong to important widespread universality classes: bulk limit leads to equations for the Sine kernel,$^{13,17}$ whereas edge limit (largest eigenvalues) leads to Airy kernel.$^{16,17}$ One studies the operator $K_n^J$ (4) with the kernel (5), (6) and $b_{n-1} = \sqrt{n/2}$ for the Hermite case. Also important for this theory is the resolvent kernel of $K$ (we will use short-hand notation $K$ for $K_n^J$), $R(x, y)$, the kernel of $K(I-K)^{-1}$, which is defined by the operator identity

$$
(I + R)(I - K) = I.\ \eqno(35)
$$

Then one introduces$^{17}$ the auxiliary functions (denoted in Ref. $17$ as $Q$ and $P$, respectively)

$$
q(x;J) = (I-K)^{-1}\phi(x), \hspace{1cm} p(x;J) = (I-K)^{-1}\psi(x),\ \eqno(36)%(4-4)
$$

\noindent and auxiliary inner products, which are functions of only the endpoints $a_k$ of $J$,

$$
u = (q, \phi\chi_{J^c}), \hspace{1cm} v = (q, \psi\chi_{J^c}) = (p, \phi\chi_{J^c}), \hspace{1cm} w = (p, \psi\chi_{J^c}).\ \eqno(37)%(4-5)
$$

\noindent Then the resolvent kernel $R(x, y)$ is$^{12,17}$

$$
R(x, y) = \frac{q(x;J)p(y;J) - p(x;J)q(y;J)}{x-y},\ (x,y \in J^c,\ x \neq y)\ \eqno(38)%(4-6)
$$

\noindent and $R(x, x) = p(x;J)q'(x;J) - q(x;J)p'(x;J)$. The TW equations for 1-matrix case are the equations for the functions of the endpoints --- $R(a_j, a_k)$,

$$
q_k = q(a_k;J), \hspace{1cm} p_k = p(a_k;J),
$$

\noindent and the auxiliary functions $u$, $v$, $w$. Among them there are universal equations valid for any Hermitian 1-matrix unitarily invariant model. They read$^{17}$

$$
\frac{\prt q_j}{\prt a_k} = (-1)^kR(a_j, a_k)q_k,\ \eqno(39)%(4-7)
$$

$$
\frac{\prt p_j}{\prt a_k} = (-1)^kR(a_j, a_k)p_k,\ \eqno(40)%(4-8)
$$

$$
R_{jk} \equiv R(a_j, a_k) = \frac{q_jp_k - p_jq_k}{a_j-a_k},\ \eqno(41)%(4-9)
$$

\noindent for $j \ne k$ and

$$
R_k \equiv R(a_k, a_k) = p_k\frac{\prt q_k}{\prt a_k} - q_k\frac{\prt p_k}{\prt a_k},\ \eqno(42)%(4-10)
$$

$$
\frac{\prt u}{\prt a_k} = (-1)^kq_k^2,\ \eqno(43)%(4-11)
$$

$$
\frac{\prt v}{\prt a_k} = (-1)^kq_kp_k,\ \eqno(44)%(4-12)
$$

$$
\frac{\prt w}{\prt a_k} = (-1)^kp_k^2.\ \eqno(45)%(4-13)
$$

\noindent The other equations are not universal, their particular form depends on the potential $V(x)$. For the finite $n$ Gaussian (Hermite) case they are$^{17}$

$$
\frac{\prt q_j}{\prt a_j} = -a_jq_j + (\sqrt{2n}-2u)p_j - \sum_{k\ne j}(-1)^kR(a_j, a_k)q_k,\ \eqno(46)%(4-14)
$$

$$
\frac{\prt p_j}{\prt a_j} = a_jp_j - (\sqrt{2n}+2w)q_j - \sum_{k\ne j}(-1)^kR(a_j, a_k)p_k,\ \eqno(47)%(4-15)
$$

\noindent along with

$$
R(a_j, a_j) = -2a_jq_jp_j + (\sqrt{2n}-2u)p_j^2 + (\sqrt{2n}+2w)q_j^2 + \sum_{k\ne j}(-1)^kR(a_j, a_k)(q_jp_k - p_jq_k),\ \eqno(48)%(4-16)
$$

$$
\frac{\prt R(a_j, a_j)}{\prt a_j} = -2q_jp_j - \sum_{k\ne j}(-1)^kR(a_j, a_k)^2.\ \eqno(49)%(4-17)
$$

\par It follows from (39)--(41), (46) and (47) that

$$
\sum_k\frac{\prt}{\prt a_k}q_jp_j = (\sqrt{2n}-2u)p_j^2 - (\sqrt{2n}+2w)q_j^2,
$$

\noindent and the last equation together with (43), (45) gives the first integral obtained in Ref. $17$:

$$
2\sum_j(-1)^jq_jp_j = -(2u-\sqrt{2n})(2w+\sqrt{2n}) - 2n,
$$

\noindent or, if we denote $\sqrt{2n}-2u = 2{\tilde Q_n}$, $2w+\sqrt{2n} = 2{\tilde P_n}$, as in formula (8), we get

$$
\sum_j(-1)^jq_jp_j = 2{\tilde Q_n}{\tilde P_n} - n,\ \eqno(50)%(4-18)
$$

\noindent From (39)--(41), (46) and (47) it follows also that

$$
\sum_k\frac{\prt q_j}{\prt a_k} = -a_jq_j + (\sqrt{2n}-2u)p_j,\ \eqno(51)%(4-19)
$$

$$
\sum_k\frac{\prt p_j}{\prt a_k} = a_jp_j - (\sqrt{2n}+2w)q_j.\ \eqno(52)%(4-20)
$$

\noindent Then consider the expression (using (43), (51), (8) and (50))

$$
\B_{-1}^2 {\tilde Q_n} = -\sum_{l,k}\frac{\prt}{\prt a_l}\frac{\prt}{\prt a_k}u = -2\sum_k(-1)^kq_k\left(\sum_l\frac{\prt q_k}{\prt a_l}\right) = 2\sum_k(-1)^kq_k(a_kq_k - 2{\tilde Q_n}p_k) 
$$

$$
= 2\left(\sum_ka_k(-1)^kq_k^2 - 2{\tilde Q_n}\sum_k(-1)^kq_kp_k\right) = -2\sum_ka_k\frac{\prt {\tilde Q_n}}{\prt a_k} -4{\tilde Q_n}(2{\tilde Q_n}{\tilde P_n} - n),
$$

\noindent which is nothing but

$$
\B_{-1}^2 {\tilde Q_n} = 2(-\B_0 + 2n){\tilde Q_n} - 8{\tilde Q_n}^2{\tilde P_n},\ \eqno(2')%(+)
$$

\noindent in terms of the earlier introduced ${\tilde Q_n}$. Completely analogously one can obtain

$$
\B_{-1}^2 {\tilde P_n} = 2(\B_0 + 2n){\tilde P_n} - 8{\tilde Q_n}{\tilde P_n}^2,\ \eqno(3')%(-)
$$

\noindent in terms of ${\tilde P_n}$ introduced in (8).

\par Now let us prove the last equation

$$
\B_{-1}^2 \ln\tau_n^J = 2(2{\tilde Q_n}{\tilde P_n} - n) \ \eqno(1')%(0)
$$

\noindent with the same $\tilde Q_n$ and $\tilde P_n$ defined by (8). Recall formula (4) for the Fredholm determinant as a ratio of $\tau$-functions. On the one hand, it follows from (4) that

$$
\B_{-1}\ln\tau_n^J = \sum_k\frac{\prt}{\prt a_k}\ln\tau_n^J = \sum_k\frac{\prt}{\prt a_k}\ln\det(I-K_n^J),\ \eqno(53)%(4-23)
$$

\noindent on the other hand,

$$
\sum_k\frac{\prt}{\prt a_k}\ln\det(I-K_n^J) = \sum_k\frac{\prt}{\prt a_k}Tr\ln(I-K_n^J) = 
$$

$$
= Tr(I-K_n^J)^{-1}\sum_k\frac{\prt}{\prt a_k}K_n(x,y)\chi_{J^c}(y;a) = -\sum_k(-1)^k R_k.\ \eqno(54)%(4-24)
$$

\noindent Thus, we have

$$
\B_{-1}\ln\tau_n^J = -\sum_k(-1)^k R_k.\ \eqno(55)%(4-25)
$$

\noindent Taking expression (48) for $R_k \equiv R(a_k,a_k)$ we get

$$
\sum_k(-1)^k R_k = -2\sum_k(-1)^k (a_kq_kp_k - {\tilde Q_n}p_k^2 - {\tilde P_n}q_k^2) - \sum_{j,k\ne j}(-1)^{j+k} R_{jk}(p_jq_k-q_jp_k),
$$

\noindent the last double sum above, however, is equal to zero. By (43), (45), the last expression is

$$
\sum_k(-1)^k R_k = -2\sum_k(-1)^k a_kq_kp_k + 2{\tilde Q_n}\sum_k\frac{\prt {\tilde P_n}}{\prt a_k} - 2{\tilde P_n}\sum_k\frac{\prt {\tilde Q_n}}{\prt a_k}.\ \eqno(56)%(4-26)
$$

\noindent Another auxiliary formula we need follows from (51), (52), (43) and (45):

$$
\sum_{k,j}(-1)^ka_k\frac{\prt (q_kp_k)}{\prt a_j} = 2\sum_k(-1)^ka_k({\tilde Q_n}p_k^2-{\tilde P_n}q_k^2) = 2\sum_ka_k \left({\tilde Q_n}\frac{\prt {\tilde P_n}}{\prt a_k} + {\tilde P_n}\frac{\prt {\tilde Q_n}}{\prt a_k}\right) = 2\B_0({\tilde Q_n}{\tilde P_n}).\ \eqno(57)%(4-27)
$$

\noindent Using (56),(57) and (50), we derive from (55):

$$
\B_{-1}^2\ln\tau_n^J = 2\sum_k(-1)^ka_k\sum_j\frac{\prt (q_kp_k)}{\prt a_j} + 2\sum_k(-1)^k q_kp_k - 2{\tilde Q_n}\B_{-1}^2{\tilde P_n} + 2{\tilde P_n}\B_{-1}^2{\tilde Q_n} = 
$$

$$
= 4\B_0({\tilde Q_n}{\tilde P_n}) + 4{\tilde Q_n}{\tilde P_n} - 2n - 2{\tilde Q_n}\B_{-1}^2{\tilde P_n} + 2{\tilde P_n}\B_{-1}^2{\tilde Q_n},
$$

\noindent and, using (2$'$), (3$'$), this is exactly

$$
\B_{-1}^2\ln\tau_n^J = 2(2\tilde Q_n\tilde P_n - n).
$$

\noindent Equation (1$'$) with TW variables is proven. Thus, we have shown that Toda lattice stands behind TW derivations. 

\section*{\normalsize\bf V. TW VARIABLES IN TERMS OF $\tau$-FUNCTIONS}

\par The TW $u$ and $w$ variables for Gaussian case turn out to be directly related to the ratios of Toda lattice $\tau$-functions $\tau_{n+1}/\tau_n$ and $\tau_{n-1}/\tau_n$. Moreover, if we compare formulas (1$'$), (44) and (50), we see that in the Gaussian case, the equation holds:

$$
\B_{-1}^2\ln\tau_n^J = 2(2\tilde Q_n\tilde P_n - n) = 2\B_{-1}v,\ \eqno(58)%(5-0)
$$

\noindent or

$$
\B_{-1}(\B_{-1}\ln\tau_n^J - 2v) = 0.\ \eqno(59)%(5-1)
$$

\noindent Besides, the equations (58) and (55), (56) mean, respectively, also the following relations among $\tau$-functions and TW variables in the Gaussian case:

$$
Q_nP_n = {\tilde Q_n}{\tilde P_n},\ \eqno(60)%(5-2)
$$

$$
\B_{-1}\ln\tau_n^J = 2\B_{0}v + 2{\tilde P_n}\B_{-1}{\tilde Q_n} - 2{\tilde Q_n}\B_{-1}{\tilde P_n}.\ \eqno(61)%(5-3)
$$

\noindent Then from (2), (3) and (60) it follows immediately that

$$
\frac{\B_{-1}^2 Q_n + 2\B_0Q_n}{Q_n} = \frac{\B_{-1}^2 \tilde Q_n + 2\B_0\tilde Q_n}{\tilde Q_n} = \frac{\B_{-1}^2 P_n - 2\B_0P_n}{P_n} = \frac{\B_{-1}^2 \tilde P_n - 2\B_0\tilde P_n}{\tilde P_n} = 4(n - 2Q_nP_n).\ \eqno(62)%(5-4)
$$

\noindent From equations (62) one can readily obtain by simple algebra:

$$
\B_{-1}(P_n\B_{-1}Q_n - Q_n\B_{-1}P_n) = -2\B_0(Q_nP_n)\ \eqno(63)%(5-5)
$$

\noindent as well as the identical equation for the corresponding ``tilde"-quantities. Since the right-hand side of equation (63) is equal to the one for ``tilde"-quantities, the corresponding left-hand sides are equal too. Now we can apply the same argument as was used in Ref. $17$ to obtain the crucial first integral (50) in this finite $n$ Hermite (Gaussian) case, this time for the function

$$
F_n = P_n\B_{-1}Q_n - Q_n\B_{-1}P_n = \tilde P_n\B_{-1}\tilde Q_n - \tilde Q_n\B_{-1}\tilde P_n.\ \eqno(64)%(5-6)
$$

\noindent The argument consists in that the last equality in (64) is true because clearly both sides tend to zero as all their arguments $a_i$ tend to $\infty$, and by (63) and (62) both are translationally invariant w.r.t. any vector of the form $(a, a, ..., a)$. Then by (62) we can rewrite (64) as

$$
\B_{-1}\ln \tilde Q_n - \B_{-1}\ln \tilde P_n = \B_{-1}\ln Q_n - \B_{-1}\ln P_n \ \eqno(65)%(5-7)
$$

\noindent Besides, by (62) we also have

$$
\B_{-1}\ln \tilde Q_n + \B_{-1}\ln \tilde P_n = \B_{-1}\ln Q_n + \B_{-1}\ln P_n,
$$

\noindent and so the simple relations emerge,

$$
\B_{-1}\ln \tilde Q_n = \B_{-1}\ln Q_n,\ \eqno(66)%(*+)
$$

$$
\B_{-1}\ln \tilde P_n = \B_{-1}\ln Q_n.\ \eqno(67)%(*-)
$$

\noindent Now by applying the above argument$^{17}$ once again, the announced result $\tilde Q_n = C_nQ_n$, $\tilde P_n = Q_n/C_n$ with some constant $C_n$ follows. The constant $C_n$ is possible to determine from the recursion, by definition of $\tau$-ratios $Q_n$, $P_n$:

$$
P_n = \frac{1}{Q_{n-1}}
$$

\noindent and the limit when $J$ becomes the whole real line. In this limit both $u$ and $w$ tend to zero, and the limit of $\tilde Q_n\tilde P_n$ is just $n/2$. Then also $Q_n/Q_{n-1} = n/2$ (this is as it should be because $Q_nP_n = b_{n-1}^2$ in this limit) and, comparing with the well-known value of the size $n$ GUE matrix integral over the whole space $\reals^n$ (see, e.g., Ref. $14$),

$$
\tau_n(0) = \frac{\pi^{n/2}}{2^{n(n-1)/2}}\prod_{j=1}^{n-1}j!,\ \eqno(68)
$$

\noindent we find that 

$$
C_n = \frac{2^n}{n!}\sqrt\frac{n}{2\pi}.\ \eqno(69)
$$

On the other hand, equation (58) is also possible to integrate once by the same argument that in the limit of all $a_i \to \infty$ both functions $v$ and $B_{-1}\ln\tau_n$ tend to zero. So by translation invariance they must be equal:

$$
v = \B_{-1}\ln\tau_n^J/2.\ \eqno(7)%(*0)
$$

\section*{\normalsize\bf VI. CONCLUSION}

\par A clear connection is exposed between the structures of equations derived by different --- Tracy-Widom and Adler-Shiota-van Moerbeke --- methods. The structure of universal equations arising from resolvent kernel analysis is identified with that of 1-Toda--AKNS integrable hierarchy. We have shown how, in the Gaussian case, to express the auxiliary coordinates, related to the resolvent kernel in the first approach, in terms of ratios of $\tau$-functions, the natural coordinates of the second approach. The last ratios have here also a simple meaning of probabilities. The close relation between Toda lattice hierarchies and systems of orthogonal functions is crucial for this connection.  
\par Although our explicit results have been obtained for single Gaussian matrix case, everything we have done here can as well be implemented for other one-matrix Hermitian unitarily invariant ensembles. Also it is well-known that the (more universal) equations for various large $n$ limits can be readily obtained from the ones for finite $n$ (see, e.g., Ref. $17$ or Ref. $3$). Moreover, our results pave the way to finding some similar connections for the case of coupled Hermitian matrices. Such relations then may facilitate the more difficult analysis of probabilities for the spectrum of coupled random matrices, with many possible applications to random growth processes and other nonstationary nonlinear dynamics phenomena. This work is in progress. 

\bigskip
\bigskip

\noindent {\bf ACKNOWLEDGMENTS} \\
\par The author would like to express particular gratitude to his scientific adviser C.A.Tracy, without whose sharing of expertise and constant encouragement and support this work could not be done. The author wishes to acknowledge the CRM lectures by M.Adler and P.van Moerbeke which were most helpful in understanding their work; to thank A.Yu.Orlov and A.Soshnikov for useful discussions; J.Harnad and J.Hurtubise for organizing the CRM Summer School on ``Random Matrices, Random Processes and Integrable Systems", Montr\'eal, June--July 2005, and for hospitality. Research and travel was supported in part by National Science Foundation through grant DMS-0553379 and VIGRE grant DMS-0135345, and by CRM, Montr\'eal, Canada.

\section*{\normalsize\bf VII. APPENDIX: J.HARNAD'S PROCEDURE -- FINITE $n$ GAUSSIAN CASE}

\par Following notations of Harnad,$^{11}$ introduce new TW variables:

$$
X_{2j} = 2iq_{2j},\ X_{2j+1} = 2q_{2j+1},\ \eqno(A1)
$$

$$
Y_{2j} = ip_{2j},\ Y_{2j+1} = p_{2j+1}.\ \eqno(A2)
$$

\noindent Then, in terms of these variables, the first derivatives of $\tau$-functions are the Poisson commuting Hamiltonians $G_j^H$ for a finite integrable dynamical system$^{11}$ with canonical coordinates $\{X_j,Y_j,u,w\}$ in our case,

$$
G_j^H = \frac{\prt \ln\tau_n^J}{\prt a_j} = (-1)^{j+1}R(a_j,a_j) = 
$$

$$
= -a_jX_jY_j + (\sqrt{2n}-2u)Y_j^2 + 1/4(\sqrt{2n}+2w)X_j^2 - 1/4\sum_{k\neq j}^{2m}\frac{(X_jY_k-Y_jX_k)^2}{a_j-a_k},\ \eqno(A3)
$$

\noindent the relevant TW equations are

$$
\frac{\prt X_j}{\prt a_k} = -1/2\frac{X_jY_k-Y_jX_k}{a_j-a_k}X_k,\ \eqno(39')%(4.7)
$$

$$
\frac{\prt Y_j}{\prt a_k} = -1/2\frac{X_jY_k-Y_jX_k}{a_j-a_k}Y_k,\ \eqno(40')%(4.8)
$$

\noindent for $j \ne k$ and

$$
\frac{\prt u}{\prt a_k} = -1/4X_k^2,\ \eqno(43')%(4.11)
$$

$$
\frac{\prt w}{\prt a_k} = -Y_k^2,\ \eqno(45')%(4.13)
$$

$$
\frac{\prt X_j}{\prt a_j} = -a_jX_j + 2(\sqrt{2n}-2u)Y_j + 1/2\sum_{k\ne j}^{2m}\frac{X_jY_k-Y_jX_k}{a_j-a_k}X_k,\ \eqno(46)%(4.14)
$$

$$
\frac{\prt Y_j}{\prt a_j} = a_jY_j - 1/2(\sqrt{2n}+2w)X_j + 1/2\sum_{k\ne j}^{2m}\frac{X_jY_k-Y_jX_k}{a_j-a_k}Y_k.\ \eqno(47)%(4.15)
$$

\noindent Let 

$$
R_k^H = \sum_{j=1}^{2m}a_j^kG_j^H,\ \eqno(A4)
$$

$$
Q_k = \sum_{j=1}^{2m}a_j^kX_j^2,\ P_k = \sum_{j=1}^{2m}a_j^kY_j^2,\ S_k = \sum_{j=1}^{2m}a_j^kX_jY_j.\ \eqno(A5)
$$

\noindent Then the first integral (50) is

$$
2\sqrt{2n}(u-w) + 4uw - S_0 = 0.\ \eqno(A6)
$$

\noindent We express by the above formulas (let ${\tilde u} = \sqrt{2n}-2u$, ${\tilde w} = \sqrt{2n}+2w$):

$$
\B_{-1}\ln\tau_n^J = R_0^H = \sum_{j=1}^{2m}G_j^H = -S_1 + {\tilde u}P_0 + 1/4{\tilde w}Q_0,\ \eqno(A7)
$$

$$
\B_0\ln\tau_n^J = R_1^H = \sum_{j=1}^{2m}a_jG_j^H = -S_2 + {\tilde u}P_1 + 1/4{\tilde w}Q_1 - 1/8\sum_{j,k}(X_jY_k-Y_jX_k)^2 =
$$

$$
= -S_2 + {\tilde u}P_1 + 1/4{\tilde w}Q_1 - 1/4Q_0P_0 + 1/4S_0^2,\ \eqno(A8)
$$

\noindent we need also

$$
\B_1\ln\tau_n^J = R_2^H = \sum_{j=1}^{2m}a_j^2G_j^H = -S_3 + {\tilde u}P_2 + 1/4{\tilde w}Q_2 - 1/8\sum_{j,k}(a_j+a_k)(X_jY_k-Y_jX_k)^2 =
$$

$$
= -S_3 + {\tilde u}P_2 + 1/4{\tilde w}Q_2 - 1/4(Q_0P_1 + Q_1P_0) + 1/2S_0S_1,\ \eqno(A9)
$$

\noindent because $\sum_{j,k}(a_j+a_k)(X_jY_k-Y_jX_k)^2 = 2\sum_{j,k}a_j(X_j^2Y_k^2+Y_j^2X_k^2-2X_jX_kY_jY_k) = 2(Q_1P_0+Q_0P_1-2S_0S_1)$. We now find from (A7)--(A9) all the terms appearing in the ASvM boundary-KP equation (34). Applying the operators $\B_k$ to (A7)--(A9), we use the Poisson commutativity$^{11}$ of $\{G_j^H\}_{j=1,...,2m}$. This property means that the derivatives of $R_j^H$ with respect to the $a_j$'s are just given by their {\it explicit} dependence on these parameters.$^{11}$ This greatly simplifies the necessary differentiations so we obtain:

$$
\B_{-1}^2\ln\tau_n^J = \B_{-1}R_0^H = -S_0,\ \eqno(A10)
$$

$$
\B_{-1}\B_1\ln\tau_n^J = \B_{-1}R_2^H = -3S_2 + 2{\tilde u}P_1 + 1/2{\tilde w}Q_1 - 1/2Q_0P_0 + 1/2S_0^2,\ \eqno(A11)
$$

$$
\B_0^2\ln\tau_n^J = \B_0R_1^H = -2S_2 + {\tilde u}P_1 + 1/4{\tilde w}Q_1,\ \eqno(A12)
$$

$$
\B_{-1}^4\ln\tau_n^J = \B_{-1}^2(\B_{-1}^2\ln\tau_n^J) = \B_{-1}^2(B_{-1}R_0^H) = -\B_{-1}^2S_0 = -\sum_i\frac{\prt}{\prt a_i}\sum_l\frac{\prt}{\prt a_l}\sum_jX_jY_j =
$$

$$
= \sum_i\frac{\prt}{\prt a_i}\sum_j[2{\tilde u}Y_j^2 - 1/2{\tilde w}X_j^2] =
$$

$$
= -\sum_j[2{\tilde u}\cdot2Y_j(a_jY_j-{\tilde w}X_j/2) - {\tilde w}/2\cdot2X_j(-a_jX_j + 2{\tilde u}Y_j)] - \sum_j[4Y_j^2\cdot\sum_iX_i^2/4 + X_j^2\cdot\sum_iY_i^2]
$$

$$
= \sum_j[-2\sum_iX_i^2Y_j^2 - 4a_j{\tilde u}Y_j^2 - a_j{\tilde w}X_j^2 + 4{\tilde u}{\tilde w}X_jY_j] =
$$

$$
= -2Q_0P_0 - 4{\tilde u}P_1 - {\tilde w}Q_1 + 4{\tilde u}{\tilde w}S_0.\ \eqno(A13)
$$

\noindent Now we substitute the expressions (A8), (A10)--(A13) into the left-hand side of equation (34) and obtain

$$
(\B_{-1}^4 + 8n\B_{-1}^2 + 12\B_{0}^2 + 24\B_{0} - 16\B_{-1}\B_{1})\ln\tau_n + 6(\B_{-1}^2\ln\tau_n)^2 = 
$$

$$
= [-2Q_0P_0 - 4{\tilde u}P_1 - {\tilde w}Q_1 + 4{\tilde u}{\tilde w}S_0] - 8nS_0 + 12[-2S_2 + {\tilde u}P_1 + 1/4{\tilde w}Q_1] + 
$$

$$
24[-S_2 + {\tilde u}P_1 + 1/4{\tilde w}Q_1 - 1/4Q_0P_0 + 1/4S_0^2] -16[-3S_2 + 2{\tilde u}P_1 + 1/2{\tilde w}Q_1 - 1/2Q_0P_0 + 1/2S_0^2] + 6S_0^2
$$

$$
= 4S_0[(\sqrt{2n}-2u)(\sqrt{2n}+2w) - 2n + S_0] = 4S_0[-2\sqrt{2n}(u-w) - 4uw + S_0] = 0,
$$

\noindent by the TW first integral (A6). Thus, the ASvM boundary-KP equation for the finite $n$ Gaussian case (34) is proved by expressing the derivatives of the $\tau_n^J$-function in terms of the TW variables, in the way of Ref. $11$.

%\bibliography{TWandAvMjmp}

%\begin{thebibliography}{10}
\bigskip

%\begin{description}%{enumerate}[itemsep=0pt,parsep=0pt,leftmargin=*,labelsep=0pt]%label=$^{\arabic*}$,

{\small

\noindent $^1$\parbox[t]{\textwidth}{M.~Adler and P.~van~Moerbeke,%\bibitem{AvM2}
\newblock ``Matrix integrals, Toda symmetries, Virasoro constraints and
  orthogonal polynomials",
\newblock Duke Math. Journal {\bf 80}, 863--911, (1995).} \\
$^2$\parbox[t]{\textwidth}{M.~Adler, T.~Shiota and P.~van~Moerbeke,%\bibitem{AvM5}
\newblock ``Random matrices, vertex operators and the Virasoro algebra",
\newblock Physics Letters A {\bf 208}, 67--78, (1995).} \\
$^3$\parbox[t]{\textwidth}{M.~Adler and P.~van~Moerbeke,%\bibitem{AvM7}
\newblock ``Hermitian, symmetric and symplectic random ensembles: PDEs for the distribution of the spectrum",
\newblock Ann. Math. {\bf 153}, 149--189, (2001).} \\
$^4$\parbox[t]{\textwidth}{M.~Adler and P.~van~Moerbeke,%\bibitem{AvM1}
\newblock ``The spectrum of coupled random matrices",
\newblock Ann. Math. {\bf 149}, 921--976, (1999).} \\
$^5$\parbox[t]{\textwidth}{M.~Adler and P.~van~Moerbeke,%\bibitem{AvM3}
\newblock ``PDEs for the joint distributions of the Dyson, Airy and Sine
  processes",
\newblock Ann. Prob. {\bf 33}, 1326--1361, (2005).} \\
$^6$\parbox[t]{\textwidth}{M.~Adler and P.~van~Moerbeke,%\bibitem{AvMdKP}
\newblock ``Generalized orthogonal polynomials, discrete KP and Riemann-Hilbert problems",
\newblock Commun. Math. Phys. {\bf 207}, 589--620, (1999).} \\
$^7$\parbox[t]{\textwidth}{H.~Awata, Y.~Matsuo, S.~Odake and J.~Shiraishi,%\bibitem{AwataEtAl}
\newblock ``Collective Field Theory, Calogero-Sutherland Model and Generalized Matrix Models",
\newblock Phys. Lett. B {\bf 347}, 49--55, (1995).} \\
$^8$\parbox[t]{\textwidth}{M.~Bertola, B.~Eynard and J.~Harnad,%\bibitem{BeEyHa}
\newblock ``Duality, biorthogonal polynomials and multi–matrix models",
\newblock Commun. Math. Phys. {\bf 229}, 73--120, (2002).} \\
$^9$\parbox[t]{\textwidth}{E.~Date, M.~Jimbo, M.~Kashiwara and T.~Miwa,%\bibitem{DJKM}
\newblock ``Transformation Groups for Soliton Equations",
\newblock in {\em Non-linear integrable systems -- classical theory and quantum theory, Proceedings of RIMS Symposium}, 39--119, (World Scientific, Singapore, 1983).} \\
$^{10}$\parbox[t]{\textwidth}{L.~Haine and J.-P.~Semengue,%\bibitem{HaSe}
\newblock ``The Jacobi polynomial ensemble and the Painlev\'e VI equation",
\newblock Journal of Mathematical Physics {\bf 40}, 2117--2134, (1999).} \\
$^{11}$\parbox[t]{\textwidth}{J.~Harnad,%\bibitem{Har}
\newblock ``On the bilinear equations for Fredholm determinants appearing in
  random matrices",
\newblock Journal of Nonlinear Mathematical Physics {\bf 9}, 530--550, (2002).} \\
$^{12}$\parbox[t]{\textwidth}{A.~Its, A.~Izergin, V.~Korepin and N.~Slavnov,%\bibitem{IIKS-90}
\newblock ``Differential equations for quantum correlation functions",
\newblock International Journal of Modern Physics B {\bf 4}, 1003--1037, (1990).} \\
$^{13}$\parbox[t]{\textwidth}{M.~Jimbo, T.~Miwa, Y.~M\^ori and M.~Sato,%\bibitem{JMMS}
\newblock ``Density matrix of an impenetrable Bose gas and the fifth Painlev\'e transcendent",
\newblock Physica D {\bf 1}, 80--158, (1980).} \\
$^{14}$\parbox[t]{\textwidth}{M.-L.~Mehta,%\bibitem{Me04}
\newblock {\em Random matrices, 3rd ed.}
\newblock (Elsevier, San Diego, 2004).} \\
$^{15}$\parbox[t]{\textwidth}{A.~Newell,%\bibitem{Newell}
\newblock {\em Solitons in Mathematics and Physics}
\newblock (Cambridge University Press, 1985).} \\
$^{16}$\parbox[t]{\textwidth}{C.A.~Tracy and H.~Widom,%\bibitem{TW-Airy}
\newblock ``Level-Spacing Distributions and the Airy Kernel",
\newblock Commun. Math. Phys. {\bf 159}, 151--174, (1994).} \\
$^{17}$\parbox[t]{\textwidth}{C.A.~Tracy and H.~Widom,%\bibitem{TW1}
\newblock ``Fredholm determinants, differential equations and matrix models",
\newblock Commun. Math. Phys. {\bf 163}, 38--72, (1994).} \\
$^{18}$\parbox[t]{\textwidth}{C.A.~Tracy and H.~Widom,%\bibitem{TW2}
\newblock ``Differential equations for Dyson processes",
\newblock Commun. Math. Phys. {\bf 252}, 7--41, (2004).} \\
$^{19}$\parbox[t]{\textwidth}{K.~Ueno and K.~Takasaki,%\bibitem{UT84}
\newblock ``The Toda Lattice Hierarchy",
\newblock Adv. Stud. Pure Math. {\bf 4}, 1--95, (1984).}

}

%\end{description}%{enumerate}
%\end{thebibliography}

\end{document}